# Contribution of Coincidence Detection to Speech Segregation in Noisy Environments

Asaf Zorea [*1] and Miriam Furst [†2]

[1]Department of Electrical Engineering, Tel-Aviv University
[2]Department of Electrical Engineering, Tel-Aviv University

**Abstract**

This study introduces a biologically-inspired model designed to examine the role of coincidence detection cells in speech segregation tasks. The model consists of three stages: a time-domain cochlear model that generates instantaneous rates of auditory nerve fibers, coincidence detection cells that amplify neural activity synchronously with speech presence, and an optimal spectro-temporal speech presence estimator. A comparative analysis between speech estimation based on the firing rates of auditory nerve fibers and those of coincidence detection cells indicates that the neural representation of coincidence cells significantly reduces noise components, resulting in a more distinguishable representation of speech in noise. The proposed framework demonstrates the potential of brainstem nuclei processing in enhancing auditory skills. Moreover, this approach can be further tested in other sensory systems in general and within the auditory system in particular.

## 1 Introduction

In our daily lives, following a conversation often involves listening to speech accompanied by some background noise. The auditory system adeptly processes and discriminates complex acoustic information, allowing us to extract relevant speech cues from the surrounding sound. Previous studies have demonstrated that speech segregation, the process of separating speech from noise, significantly contributes to speech perception and comprehension[1,2]. Bregman[3] ascribes auditory segregation to auditory scene analysis and outlines two stages involved in the segregation process: segmentation and grouping. During segmentation, the input is divided into segments. In the grouping stage, the segments that are estimated to originate from the same source are clustered together. Numerous studies have adopted the auditory scene analysis approach to achieve comprehensive speech segregation. A common technique involves employing a time-frequency (T-F) representation based on the speech spectrogram, utilizing a logarithmic scale of the frequency domain. Estimating the speech presence probability (SPP) relies on analyzing the statistical characteristics of both the speech and the background noise[4,5]. Moreover, thresholding is often utilized to generate the ideal binary mask of the speech[6–8].

[*]zoreasaf@gmail.com
[†]mira@eng.tau.ac.il

The cochlea decompose sounds into narrow-band signals with specific characteristic frequencies. Then, auditory information propagates via the auditory nerve through multiple auditory nuclei, including the cochlear nucleus and inferior colliculus. These centers extract and process complex acoustic features from the neural input. In the inferior colliculus, one of the common cell types is the coincidence detection (CD) cell[9]. This neuron encode information by detecting the occurrence of temporally close but spatially distributed input signals. Krips and Furst[10] have shown that if the inputs act as a non-homogeneous Poisson process (NHPP), then the CD output also behaves as NHHP. The extracted information is transmitted to the auditory cortex, which is further processed and integrated over time to contribute to the comprehension and perception of spoken language.

This study aims to investigate the potential involvement of CD neurons in speech segregation using biologically motivated computational modeling. The model presented in this study includes three key stages: In the first stage, an initial T-F representation is obtained by a cochlear model, which generates instantaneous rates (IRs) of auditory nerve fibers (ANFs)[11–14]. In the second stage, a network of CD cells is integrated to enhance the neural representation of the auditory input. Finally, an optimal speech presence estimator is employed, enabling us to assess the effectiveness of the CD processing. The structure of this paper is organized as follows. The material and methodology are presented in Section 2. The study results are presented in Section 3. Finally, the discussion and conclusions are summarized in Section 4 and Section 5.

## 2 Material and Methods

A schematic illustration of the model is depicted in Fig. 1. The diagram is divided into three blocks, each representing a component of the model. The first block represents the auditory periphery, which is responsible for the initial processing of auditory stimuli. The second block illustrates the network of CD cells designed with excitatory inputs. The third block signifies the speech estimator, which integrates input from multiple tonotopic channels to estimate the probability of speech presence. Notably, this estimator can receive input from either CD cells or ANFs responses.

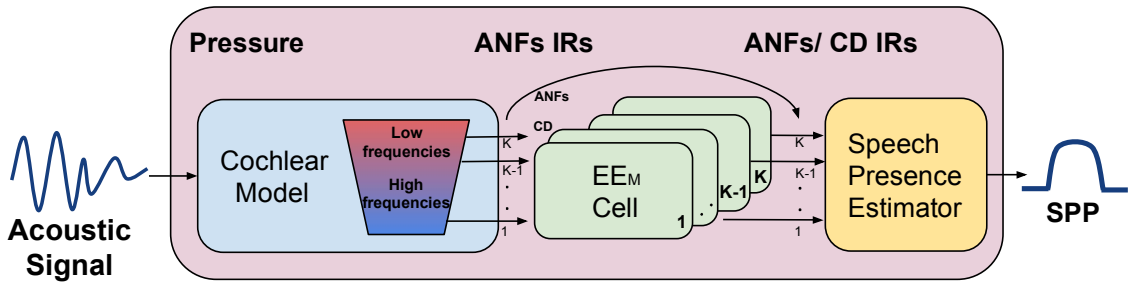

Figure 1: A schematic description of the computational model.

### 2.1 Cochlear Model

The cochlear model utilized in this study employs a time-domain solution of cochlear mechanics. It calculates the basilar membrane motion as a response to an acoustic stimulus while integrating the electro-mechanical non-linear motion of the outer hair cells[11-13,15]. Practically, the model was simulated with an adaptive time step and 256 cochlear partitions. The derivation of the ANFs' IRs at each cochlear partition was obtained by phenomenological model[14,16].

### 2.2 Coincidence Cells Architecture

Each neural input is represented by a set of spikes that occur at instances $\{t_n, n \in \mathcal{N}\}$. This series of spikes events can be described as a random point process with IR $\lambda(t)$, and refractory period $\tau_r$. A general excitatory-excitatory (EE) cell, $EE_M^N$, has $N$ independent excitatory inputs $\Psi = \{E_1, .., E_N\}$ with corresponding IRs $\Psi_\lambda = \{\lambda_{E_1}, .., \lambda_{E_N}\}$, and generates a spike when at least $M$ of its inputs spike during an interval $\Delta_c$. To maintain simplicity, it was assumed that

$M = N$ and denote it as $EE_M$. Such a cell generates spikes at instances $\{t_{n_f}, n_f \in \mathcal{N}\}$,

$$\left.\begin{array}{l} t_{n_f} = \max\{t^1{}_{n_f}, ..., t^M{}_{n_f}\} \\ if \quad \max\{t^1{}_{n_f}, ..., t^M{}_{n_f}\} - \min\{t^1{}_{n_f}, ..., t^M{}_{n_f}\} < \Delta_c \end{array}\right\} \quad (1)$$

where $\{t^1{}_{n_f}, ..., t^M{}_{n_f}\}$ denote the discrete firing times of the $M$ excitatory inputs respectively.

According to Krips and Furst[10], CD cells exhibit NHHP behavior when their inputs are also NHPP point processes. As a result, their output can be computed analytically. The expression for the $EE_M$ cell's IR was obtained using this approach:

$$\lambda_{EE_M}(t|\Psi_\lambda) = \sum_{m=1}^{M} [\lambda_{E_m}(t) \cdot \prod_{\tilde{m}=1, \tilde{m} \neq m}^{M} \int_{t-\Delta_c}^{t} \lambda_{E\tilde{m}}(t)] \quad (2)$$

Despite the diversity of the $EE_M$ cell's inputs, it is reasonable to presume that the firing rates of the $M$ neurons in response to a given stimulus would be similar on average, therefore:

$$\lambda_{E_m}(t) \triangleq \lambda_E(t), \quad \forall \ m \in \{1, .., M\} \quad (3)$$

where $m$ denotes the input cell index.

The $EE_M$ cell's output, $\lambda_{EE_M}$, may be described as follows:

$$\lambda_{EE_M}(t|\Psi_\lambda) = M \cdot \lambda_E(t) \cdot \underbrace{\left(\int_{t-\Delta_c}^{t} \lambda_E(\tau) d\tau\right)^{M-1}}_{I_c(t)} \quad (4)$$

where $I_c$ represents the coincidence integral.

A discrete $EE_M$ cell's output, $\lambda_{EE_M}[n]$, can be obtained using a discrete approximation of the coincidence integral $I_c$. For a time domain discretized into $N_c$ equal panels, each of size $\delta_s$. By applying the trapezoidal rule, an approximation for $I_c$ can be obtained by:

$$\int_{t-\Delta_c}^{t} \lambda_E(\tau) d\tau \simeq \left(\begin{array}{l} \frac{1}{2} \cdot \lambda_E(\tau_1) + \lambda_E(\tau_2) + .. + \\ + \lambda_E(\tau_{N_c-1}) + \frac{1}{2} \cdot \lambda_E(\tau_{N_c}) \end{array}\right) \cdot \delta_s \quad (5)$$

where $N_c = \lceil \Delta_c \cdot fs \rceil$ is the discrete integration window length, $\tau_i = \lfloor t \cdot f_s \rfloor + i$ the discrete time index, $\delta_s = \frac{1}{f_s}$ is the sample time, and $fs$ is the sample rate.

As a consequence, in the discrete-time domain, the coincidence integral can be computed by convolving $\lambda[n]$ with the following finite impulse response (FIR) filter $h_{fir}[n]$:

$$\left.\begin{array}{l} h_{fir}\left[n\right] = \left[\frac{1}{2}, 1, .., 1, \frac{1}{2}\right]_{N_c} \cdot \delta_s \\ I_c\left[n\right] = \lambda_E\left[n\right] * h_{fir}\left[n\right] \end{array}\right\} \quad (6)$$

Finally, the discrete $EE_M$ cell's IR, $\lambda_{EE_M}[n]$, was obtained by:

$$\lambda_{EE_M}\left[n | \Psi_\lambda\right] = M \cdot \lambda_E\left[n\right] \cdot \left(\lambda_E\left[n\right] * h_{fir}\left[n\right]\right)^{M-1} \quad (7)$$

The corresponding CD cells' IRs are generated from $K$ vectors of ANFs' IRs received.

## 2.3 Speech Presence Estimation

When an interfering noise coincides in frequency and time with a signal of interest, they both interfere on the basilar membrane, causing both the signal and the noise to compete for the same receptors. Let $\boldsymbol{\lambda_K}(n)$ be a IRs random vector distributed across $K$ cochlear partitions, as a function of time. In the neural activity domain, according to the tonotopic organization of the auditory system, it can be assumed that the neural response is an additive mixture of clean speech $\boldsymbol{\lambda}^{Speech}(n)$ and acoustic noise $\boldsymbol{\lambda}^{Noise}(n)$.

Two hypotheses $H_1[n]\,and\,H_2[n]$ were suggested, and indicate speech absence and speech presence respectively,

$$\left.\begin{array}{l} H_1\left[n\right]: \boldsymbol{Y}\left(n\right) = \boldsymbol{\lambda}^{Noise}\left[n\right] \\ H_2\left[n\right]: \boldsymbol{Y}\left(n\right) = \boldsymbol{\lambda}^{Speech}\left[n\right] + \boldsymbol{\lambda}^{Noise}\left[n\right] \end{array}\right\} \quad (8)$$

The process of separating an auditory scene into distinct objects was modeled as an unbiased optimal estimator of the SPP, which is the probability of speech being present in a noisy observation. Motivated by the central limit theorem[17], the IR's distribution, $\lambda$, was assumed to be a superposition of multivariate Gaussians generated by two parent processes:

$$p(\boldsymbol{\lambda}) = \Sigma_{i=1}^{2} \pi_i \mathcal{N}(\boldsymbol{\lambda} | \boldsymbol{\mu_i}, \boldsymbol{\Sigma_i}) \; ; \; s.t \; \sum_{i=1}^{2} \pi_i = 1 \quad (9)$$

where, correspondingly, $\mathcal{N}$ denotes a multivariate normal distribution function, $\pi_{1,2}$ denote the prior probability of $\boldsymbol{\lambda} \in H_{1,2}$, $\boldsymbol{\mu_{1,2}}$ denote the Gaussian means, and $\boldsymbol{\Sigma_{1,2}}$ denote the Gaussians covariance matrices.

Due to the statistical independence of ANFs across multiple characteristic frequencies, it was reasonable to hypothesize that any two different $\boldsymbol{\lambda}$ components are not correlated. The off-diagonal correlations were set to *zero*, resulting in a diagonal covariance matrices $\boldsymbol{\Sigma_{1,2}}$, therefore $\mathcal{N}(\boldsymbol{\lambda})$ yielded:

$$\mathcal{N}(\boldsymbol{\lambda}|\boldsymbol{\mu}, \boldsymbol{\Sigma}) = \frac{1}{(2\pi)^{K/2}} \prod_{k=1}^{K} \frac{1}{\sigma_k} \exp\left\{-\frac{1}{2}\left(\frac{\lambda_k - \mu_k}{\sigma_k}\right)^2\right\} \tag{10}$$

where $k$ and $\boldsymbol{\sigma_{1,2}}$ denote the cochlear position index and the Gaussians variances, respectively.

The problem was addressed as an optimization problem, with the objective of estimating a set of parameters that best fit the joint probability of the hypotheses, and was solved using the expectation-maximization (EM) approach[18].

Let $\boldsymbol{Z}$ be the latent vector that determine the component from $\boldsymbol{\lambda}$ originates, s.t.,

$$P(\boldsymbol{\lambda}|\boldsymbol{Z} = z) \sim \mathcal{N}(\boldsymbol{\mu}_z, \boldsymbol{\Sigma}_z) \tag{11}$$

During the expectation step, the weights $w_j[n]$ were defined as a 'soft' assignment of $\boldsymbol{\lambda}[n]$ to Gaussian $j$,

$$w_j[n] = P(z = j|\boldsymbol{\lambda[n]};\boldsymbol{\theta}) \tag{12}$$

where $\boldsymbol{\theta}$ indicates the parameters set of the model ($\boldsymbol{\theta} = \{\boldsymbol{\mu}, \boldsymbol{\sigma}, \boldsymbol{\pi}\}$).

A new parameter set $\theta$ was estimated throughout the maximization step by maximizing the log-likelihood with respect to the expectations,

$$\arg\max_\theta \sum_{n=1}^{N} \sum_{j=1}^{2} w_j[n] \log\left(\pi_j N\left(\boldsymbol{\lambda[n]}; \boldsymbol{\mu_j}, \boldsymbol{\sigma^2}_j\right)\right)\right\} \tag{13}$$

Given an initial estimate, the EM algorithm cycles through (12) to (13) repeatedly, until the estimates converge.

The entire algorithm for estimating the statistical properties of both the speech and the noise

```
Data: λ_{1,..,N}
Result: 𝒩_{j=1,2}(λ|μ_j, Σ_j)
while θ_{t+1} ≠ θ_t do
    E Step: for each n, j do
```

$$w_j[n] = \frac{\pi_j \cdot \mathcal{N}\left(\boldsymbol{\lambda}[n]|\boldsymbol{\mu}_j, \boldsymbol{\sigma}_j\right)}{\sum\limits_{j=1}^{2} \pi_j \cdot \mathcal{N}\left(\boldsymbol{\lambda}[n]|\boldsymbol{\mu}_j, \boldsymbol{\sigma}_j\right)} \qquad (14)$$

```
    end
    M Step: for each j do
```

$$\boldsymbol{\mu}_j = \frac{\sum\limits_{n=1}^{N} w_j[n] \cdot \boldsymbol{\lambda}[n]}{\sum\limits_{n=1}^{N} w_j[n]} \qquad (15)$$

$$\boldsymbol{\sigma}_j^2 = \frac{\sum\limits_{n=1}^{N} \left(\boldsymbol{\lambda}[n] - \boldsymbol{\mu}_j\right)^2 \cdot w_j[n]}{\sum\limits_{n=1}^{N} w_j[n]} \qquad (16)$$

$$\pi_j = \frac{\sum\limits_{n=1}^{N} w_j[n]}{N} \qquad (17)$$

```
    end
end
```

**Algorithm 1:** Estimating the speech presence probability using the EM algorithm with multivariate normal distribution and diagonal covariance matrix

neural activities was illustrated in Algorithm 1. After estimating all the parameters, the SPP can be obtained by:

$$SPP(\boldsymbol{\lambda}|\boldsymbol{\mu}, \boldsymbol{\sigma}) = \frac{\pi_i \mathcal{N}\left(\boldsymbol{\lambda}|\boldsymbol{\mu}_i, \boldsymbol{\sigma}_i\right)}{\sum\limits_{j=1}^{2} \pi_j \mathcal{N}\left(\boldsymbol{\lambda}|\boldsymbol{\mu}_j, \boldsymbol{\sigma}_j\right)}, \quad i \in H_2 \qquad (18)$$

## 2.4 Evaluation Method

An effective method for evaluating the ability of speech estimator to separate speech from noise is to examine the area under the receiver-operator characteristic curve (AUC), with a higher AUC indicating better performance. Threshold values in the range of $[0, 1]$ were applied to SPPs outputs to categorize them as speech presence or absent. For each threshold, the true positive rate and false positive ratio were determined by calculating the proportion of correctly identified speech-containing segments and incorrectly identified noise segments, respectively. The ground truth used for the evaluation was manually labeled by inferring which segments contain speech versus which segments contain noise.

For the evaluation, a total of thirty speech utterances were taken from the NOIZEUS database, a repository of noisy speech corpus[19]. The sentences were degraded with three different types of real-world noise: car, white, and babble. This was done through the addition of interfering signals at signal-to-noise ratios (SNRs) ranging from -15 to 15 dB, using method B of the *ITU-T P.56*[20].

## 3  Results

### 3.1  Auditory Periphary response

Fig. 2. illustrates the relationship between the cochlear response and cochlear position at different frequencies, when a linear chirp stimulus is applied at a sound pressure level (SPL) of 65 dB. The derived ANFs IRs are displayed in a color-coded format, demonstrating how the response varies with changes in input frequency along the cochlear partition.

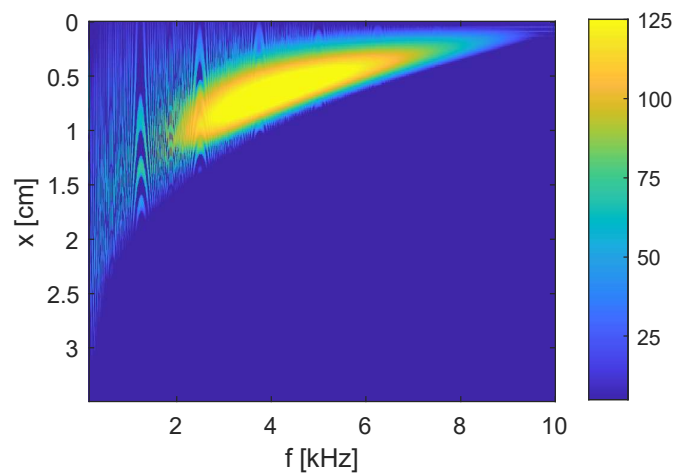

Figure 2: ANF IR derivation as a response to a linear chirp. The frequency (in kHz) is plotted along the x-axis, while the corresponding distance from the stapes (in cm) is represented on the y-axis and denoted by 'x'.

### 3.2  Example Outcome

Fig. 3. depicts an example of the model's outputs as a response to the English phrase "*We find joy in*" at level of $65$ *dB SPL*. The sentence was taken from track number $7$ of *NOIZEUS database*[19].

Fig. 3. comprises panels that depict various variables or environmental conditions. The left and right columns of the figure denoted as Panels A and B, respectively, display the model's

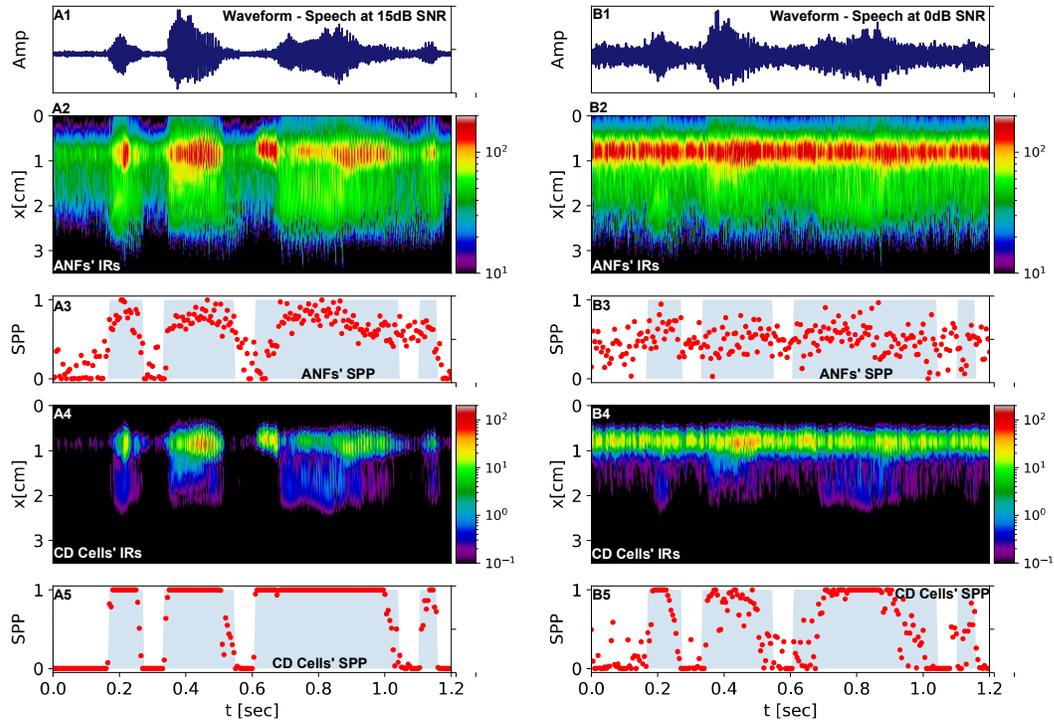

Figure 3: The acoustic waveforms, ANFs' IRs, CD cells' IRs and their corresponding SPPs were exhibited in response to the English sentence *"We find joy in"* at level of *65 dB SPL*. The sample was obtained from file 'sp07.wav' of *NOIZEUS database* between $0s$ and $1.20s$. Panels A1 and B1 respectively display the acoustic waveform for noisy speech stimuli degraded by car noise at SNRs of 0dB and 15dB. Panels A2 and B2 illustrate the ANFs' responses. Panels A3 and B3 show the corresponding ANFs' SPPs. Panels A4 and B4 display the response of the CD cells' network (with parameters $M = 6$ and $\Delta_c = 3ms$). Panels A5 and B5 provide the corresponding SPPs of the CD cells' response.

inputs and outputs for noisy speech degraded by car noise at SNRs of 0 dB and 15 dB. Panels A1 and B1 show the acoustic waveforms, while Panels A2 and B2 present the ANFs' IRs as a color-coded graph in spikes/sec, with the x-axis representing post stimulus time and the y-axis representing distance from the stapes. In Panels A3 and B3, the ANFs' SPPs are displayed with gray backgrounds indicating binary flags for speech presence (1) or absence (0). Although the SPP for speech at 15 dB SNR speech matches the manually labeled speech presence, the SPP for speech at 0 dB SNR does not clearly indicate it, regardless of the speech's presence. Panels A4 and B4 display the CD cells' IRs, while Panels A5 and B5 show their SPPs. The results show that the SPPs computed after CD processing better follow speech patterns and match manual labels, even when the energy of background noise equals that of the speech signal.

## 3.3 Coincidence Detection Cell Parameters Tuning

To determine the optimal architecture for the CD cell, we systematically varied the number of input cells ($M$) and the coincidence window ($\Delta_c$), as specified in Eq. (5). The results are presented in Figure 4. Based on these results, we selected $M = 6$ and $\Delta_c = 3ms$ as the parameters to be used in the evaluation. These parameter values correspond to those of actual CD cells found in the inferior colliculus and the ventral cochlear nucleus[21–23].

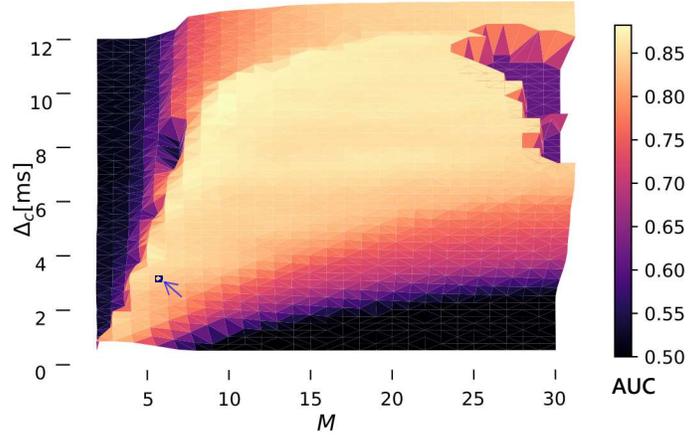

Figure 4: A color-coded graph of the AUC of speech degraded by car noise at a SNR of $0dB$, with various combinations of input cells ($M$) and coincidence window lengths ($\Delta_c$). The speech was obtained from file 'sp09.wav' of *NOIZEUS database*.

## 3.4 Speech Presence Estimators

Fig. 5 presents a comparison between CD-based and ANF-based estimators. Fig. 5a shows the noises power spectrum densities, while the average AUC scores of the 30 sentences with the corresponding standard deviations are plotted as a function of the SNR for three types of noise: babble noise (Fig. 5b), white noise (Fig. 5c), and car noise (Fig. 5d).

Both ANF-based and CD-based estimators showed an increase in average AUC with increasing SNR. However, CD-based estimators outperformed ANF-based estimators for all tested SNRs and noise types, with the most significant improvement observed for mid-low input SNRs. The statistical difference in performances was compared and yielded significant difference for all types of noises and SNRs ($P < .001$). For $SNR \geq 10\,dB$, the performance yielded by the ANF were reasonable ($AUC \geq 0.9$), thus only minor improvement was yielded by the CD processing. However, for $SNR \approx 0\,dB$ the ANF performances yielded $AUC \approx 0.7$ for all noise types, and the additional CD processing yielded $AUC \approx 0.9$. On the other hand, for very low SNRs, for example $SNR = -15\,dB$, and the ANF performances were close to chance ($AUC \approx 0.5$), the improvement yielded by the CD processing was small.

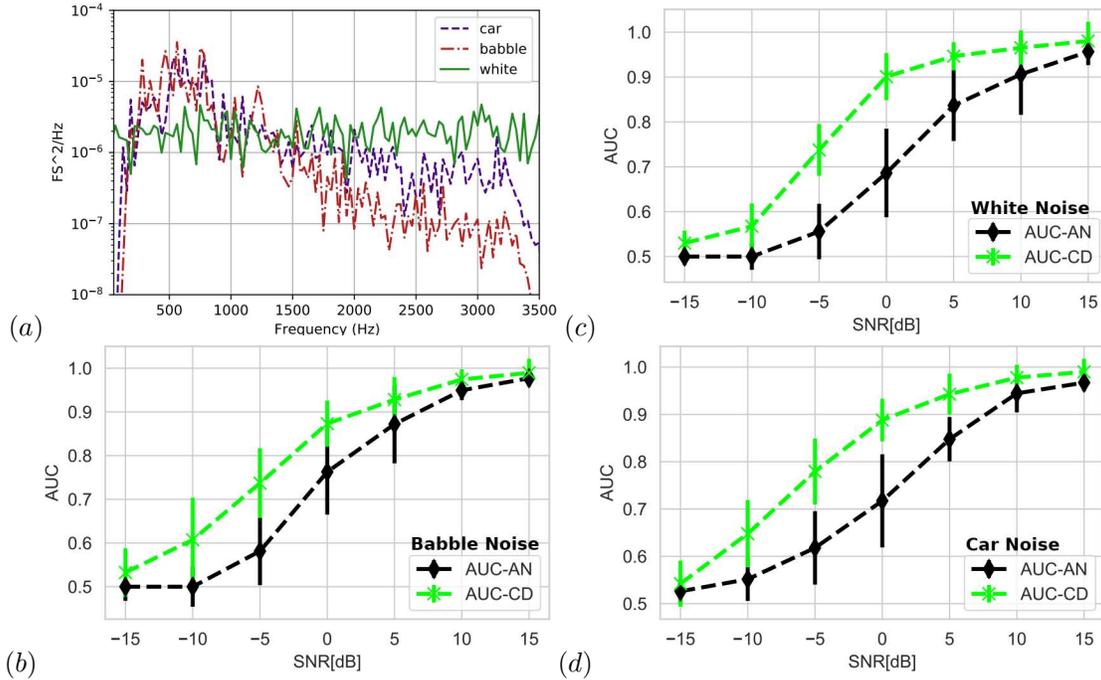

Figure 5: A comparison between ANF-based and CD-based estimators (with parameters $M = 6$, $\Delta_c = 3ms$) for a healthy cochlea. The power spectrum density and AUC scores for three different real-word noises, babble, white and car noises, at SNRs of $-15$ to $15$ dB are shown in panels $a$, $b$, $c$, and $d$ respectively.

## 4   Discussion

In this paper, a speech segregation model based on the physiology of the auditory pathway is presented. The proposed excitatory-only coincidence detection (CD) architecture demonstrates its effectiveness in reducing noise components in stationary noise while concurrently improving the accuracy of speech segregation. These findings highlight the potential of CD cells to contribute significantly to enhancing speech perception. To ensure broad applicability and avoid overfitting, the models and assumptions were simplified. Using an unsupervised optimal estimator further strengthens the study's findings, as it provides unbiased insights into the neural representation of CD processing.

CD cells are widely distributed across various auditory nuclei, with a significant presence in the trapezoid body nuclei, where they play a significant role in binaural perception[24–26]. Binaural processes have been demonstrated to enhance speech segregation[27,28], implying that CD cells may be involved in this aspect of auditory perception. However, speech segregation can also occur monaurally. In natural acoustic signals, amplitude modulation (AM) serves as a critical temporal feature, and its significance has been highlighted in various perceptual tasks, such as envelope detection and segregation[29]. Notably, CD cells have been linked to AM processing[9,30]. Furthermore, envelope and temporal fine structure information are known to be important for speech perception[31–33]. The CD cells presented in this paper function as auto-correlation units, effectively enhancing this information, which is essential for speech segregation. These findings provide valuable insights into the neural mechanisms underlying auditory processing.

While the tonotopic representation used in the estimator was found to be effective, it is important to acknowledge its limitations. The assumption of independence between different characteristic frequencies may not always hold true. Although spike generation in different auditory nerve fibers (ANFs) is statistically independent, the tuning curves of ANFs have a long low-frequency tail, and the tips of the curves broaden and decrease at higher sound pressure levels (SPLs)[34–36]. Consequently, the synaptic drive to different ANFs across the cochlear length is not entirely independent. Future investigations should incorporate more sophisticated models that account for the interactions between frequency channels. Moreover, an alternative architecture incorporating inhibitory inputs may be more effective for other types of noises or conditions. Future work should also consider including inhibitory inputs and evaluating the model's performance against different noise types.

## 5   Conclusion

Two distinct methods for speech estimation were compared: one based on coincidence detection and the other on auditory nerve fibers. CD-based estimators consistently outperformed ANF-based estimators across all tested SNRs and noise types. The improvement was most significant for mid-low input SNRs. These findings suggested that CD information plays a crucial role in speech segregation, contributing significantly to the enhanced performance of the model.

## Acknowledgments

This research was partially supported by the ISRAEL SCIENCE FOUNDATION: grant No. 563/12.